\documentclass[manuscript]{aastex}
\usepackage{emulateapj5}

\def\lesssim{\mathrel{\hbox{\rlap{\hbox{\lower3pt\hbox{$\sim$}}}{\hbox{\raise2pt\hbox{$<$}}}}}}
\def\gtrsim{\mathrel{\hbox{\rlap{\hbox{\lower3pt\hbox{$\sim$}}}{\hbox{\raise2pt\hbox{$>$}}}}}}

\shorttitle{Chandra observation of AWM~7 core}
\shortauthors{Furusho, Yamasaki, \& Ohashi}

\begin{document}

\title{Chandra observation of the core of the galaxy cluster AWM~7}

\author{T. Furusho\altaffilmark{1,2}, N. Y. Yamasaki\altaffilmark{2}, and T. Ohashi\altaffilmark{3}
}

\altaffiltext{1}{Laboratory for High Energy Astrophysics, 
NASA/GSFC, Code 662, Greenbelt, MD 20771; furusho@olegacy.gsfc.nasa.gov}

\altaffiltext{2}{Institute of Space and Astronautical Science, 3-1-1,
  Yoshinodai, Sagamihara, Kanagawa 229-8510, Japan}

\altaffiltext{3}{Department of Physics, Tokyo Metropolitan University, 
1-1 Minami-Ohsawa, Hachioji, Tokyo 192-0397, Japan}

\begin{abstract}

We present results from a {\it Chandra} observation of the core region
of the nearby X-ray bright galaxy cluster AWM~7. There are blob-like
substructures, which are seen in the energy band 2--10 keV, within 10
kpc ($20''$) of the cD galaxy NGC~1129, and the brightest sub-peak has
a spatial extent more than 4 kpc. We also notice that the central soft
X-ray peak is slightly offset from the optical center by 1 kpc.  These
structures have no correlated features in optical, infrared, or radio
band.  Energy spectrum of the hard sub-peak indicates a temperature
higher than 3 keV with a metallicity less than 0.3 solar, or a
power-law spectrum with photon index $\sim 1.2$. A hardness ratio map
and a narrow Fe-K band image jointly indicate two Fe-rich blobs
symmetrically located around the cD galaxy, with the direction
perpendicular to the sub-peak direction. In larger scales ($r<60$
kpc), the temperature gradually drops from 4 keV to 2 keV toward the
cluster center and the metal abundance rises steeply to a peak of 1.5
solar at $r \approx 7$ kpc. These results indicate that a dynamical
process is going on in the central region of AWM~7, which probably
creates heated gas blobs and drives metal injection.

\end{abstract}


\section{INTRODUCTION}

Recent {\it Chandra} and {\it XMM-Newton} data have shown with their
advanced imaging and spectroscopic capabilities rather complex
features in the cores of rich clusters, including the brightest
objects such as M87, Centaurus, and Perseus clusters. The newly
emerging properties of the cluster core are roughly summarized into three
aspects: the complex brightness structure often correlated with
radio lobes, the lack of cool gas with temperatures below 1--2 keV, and
the interesting metallicity profiles sometimes characterized by
high-metallicity rings. The {\it Chandra} images show both filamentary
structures and holes that coincide with radio lobes in some clusters.
Clear examples are the Perseus cluster \citep{fabian00,sfs02}, and
Hydra A \citep{mcnam00,david01}. The lack of the cool gas which was
expected from radiative cooling raises a problem of heat source and
heating mechanism \citep[e.g.][]{fabian01,hans02}, which are still an
unsolved question. Various possible heat sources have been proposed,
but none of the existing models seems to explain the observed thermal
structure of the cluster gas in a plausible way. The metal-abundance
profile increases simply toward the center in M~87 \citep{matsu02},
while the metallicity profiles in Perseus, Centaurus \citep{sf02} and
Abell 2199 \citep{johns02} clusters show a peak at a certain radius
from the cluster center. The precise origin of these ``high-metallicity
rings'' is not understood yet.

AWM~7 is a nearby, X-ray bright cluster at a redshift of $z=0.0173$
\citep{kg00}. The ROSAT data show a moderate cooling flow of $\sim$ 60
M$_{\odot}$ yr$^{-1}$, and a temperature drop at the center associated
with the central cD galaxy, NGC1129 \citep{nb95}. The ASCA mapping
observation revealed an isothermal intracluster medium (ICM) with a
temperature of 4 keV \citep{furu01}, and a Mpc-scale abundance
gradient from the central 0.5 solar to about 0.2 solar in the outer
region \citep{ezawa97}. These results suggest that this cluster as a
whole has been in a dynamically relaxed state. The ICM distribution,
however, is not circularly symmetric but elliptically elongated in the
east-west direction. The X-ray peak of the cD galaxy in the PSPC image
is shifted from the center of the whole cluster by 30 kpc, and from
the optical center of the cD galaxy by 3 kpc \citep{nb95,kg00}. AWM~7
therefore is thought to be still in an early stage in the cD cluster
evolution. No detection of CO emission from the cD galaxy was reported
with an upper limit for the mass of molecular gas to be $<4\times10^8
M_{\odot}$ \citep{fuji00}. The NRAO image at 1.4 GHz does not show any
significant radio emission from the cD galaxy \citep{burns80}.

In this paper, we report on the high spatial resolution X-ray image of
AWM~7, along with the temperature and abundance profiles.  This
relatively young cD cluster is the best target to look into the
processes taking place during the evolution of cD clusters and to
make comparison with other systems.

We use $H_0=50$ km s$^{-1}$ Mpc$^{-1}$ and $q_0 = 0.5$, indicating 29
kpc for $1'$ and 0.49 kpc for $1''$ at AWM~7. The solar number abundance
of Fe relative to H is taken as $4.68 \times 10^{-5}$ \citep{ag89}.

\section{OBSERVATION AND DATA REDUCTION}

The core of AWM~7 was observed with {\it Chandra} on 2000 August 19
for 47,850 s using the ACIS-I detector. The center of the cluster was
focused on the center of ACIS-I3 with an offset of $Y-3.7$ and $Z+3.7$ arcmin from the on-axis position, and we limit here our analysis to the I3 chip
only. The data were taken with the Faint mode, and the CCD temperature
was $-120^\circ$C\@. In order to check if significant flares may
affect the data, we produced a light curve for the ACIS chips I0--2
which observed the sky region with relatively higher background
contribution than the I3 chip using events in the 0.3--10 keV energy
band. We found there was no flare-like time variation.  We applied the
data screening to keep the count rates within $\pm 10 \%$ of the
average of 4.7 count s$^{-1}$, and it resulted in a slight reduction
in the usable exposure time to 47,731 s.

\section{THE CD GALAXY REGION}

\subsection{X-ray Image}
\label{sec:img}

% Figure 1a : 0.5-7 keV image
A smoothed image of the central $1'\times 1'$ region of AWM~7 in the
0.5--7 keV band is shown in Figure 1a. The image is smoothed by a
Gaussian function with $\sigma=2''$, and corrected for exposure. The
X-ray image appears to mainly consist of the diffuse ICM and three
emission components. The bright central emission corresponds to the cD
galaxy, NGC 1129. The point-like weak emission at $27''$ (13 kpc)
southwest of the center is identified as a small galaxy, VV085b, which
is more clearly seen in the 2MASS (The Two Micron All Sky Survey)
infrared image in Figure 2c.  The third component is a very faint
extended feature near the cD galaxy with a separation $13''$ (6 kpc) in
the southeast of the cD\@.  This component has no cataloged counterpart
either in NED or SIMBAD\@.  The image also shows signs of blobs or
filamentary structures around the cD galaxy, and possible small X-ray
holes at $17''$ east and $18''$ west of the center, such as those seen
in other major cD clusters more distinctly.

% Figure 1b : RGB color image, about hard sub peak 
Figure 1b shows a color-coded X-ray image of the same region, in which
X-ray photon energy is shown in three colors.  The energy assignments
for the 3 colors are 0.5--1.5 keV for red, 1.5--2.5 keV for green, and
2.5--8 keV for blue, respectively.  There is an extended yellow region
seen near the center, and the shape may be an unresolved small filament
like the one seen in the Centaurus cluster \citep{sf02}. The southeast
faint structure (hereafter hard sub-peak) in the cD galaxy is obviously
bluer compared with the central region of the cD galaxy.  X-ray color of
the galaxy VV085b is red. The temperature of this galaxy is lower than
the level of NGC 1129, however this source is too faint for a
quantitative temperature determination with spectral fits.

% Figure 2 : soft(a), hard(b), 2mass(c), projection (d)
In order to look into energy dependent structures, we create the soft
(0.5--2.0 keV) and hard (2.0--10.0 keV) band images separately as
shown in Figures 2a and 2b. The images are adaptively smoothed ith a
minimum significance of $3\sigma$. In these figures, the dotted lines
indicate the cutout region whose intensity profiles are plotted in
Figure 2d.  The images are again for the central $1'$ square region.
The hard sub-peak in the southeast region marked by a cross appears
very clearly in the hard band image, while it is hardly seen in the
soft band. The X-ray flux of this hard sub-peak is very close to the
peak at the cD galaxy, as seen in the 2--10 keV band profile (filled
circles) in Figure 2d. Since the cluster center was focused on the I3
chip center with an offset of $5'$ from the optical axis, the PSF
at the hard sub-peak is wider than the on-axis PSF\@. The spatial extent
of the sub-peak, more than $8''$, however, is wider than the PSF width
at the sub-peak position as shown by the dashed curve in Figure
2d. Therefore, this sub-peak emission is unlikely to be from a single
point source.  The distance between the sub-peak and the cD galaxy
center ($13''$ or 6 kpc) is much smaller than the isophotal radius
$r_e$ of NGC~1129: $48''$ (23 kpc) \citep{bacon85}. This sub-peak is
thought to be an internal structure of the cD galaxy, if it is not
either a foreground or a background source. It is remarkable that the
2MASS image of the cD galaxy shows no sign of a structure in the same
position (Figure 2c).

We also notice that, in the hard-band image shown in Figure 2b,
there are fainter blob-like features seen just in the northwest and
possibly a hole or filament in the west of the cD galaxy. The 1-dimensional profile in
Figure 2d in fact shows a peak at $10''$ in the opposite side of the
hard sub-peak. Furthermore, this hard band image shows a marked
deviation from the spherical symmetry. The brightness shows a fairly
sharp cutoff in the north to east direction of the cD, while it is
more diffusively extended in the west to south direction. This
morphology suggests that the hotter gas component in the central
region may be moving to the northeast direction, creating a rather
sharp edge in the front and a tailing feature behind. This rather
complex nature of the central hot gas, which is only seen in the hard
energy band ($> 2$ keV) is an entirely new finding in the present
analysis. Note that the boxes in Figure 2b show regions where spectra were extracted as described in \S\ref{sec:cDspec}.

We find another small structure in the very center of the cD
galaxy. The X-ray peak in the 0.5--2 keV band is shown by a filled
triangle in Figure 2a-c. We derived the peak position from the true
data.  The soft peak position, (R.A., decl.)$_{\rm J2000}$ = ($2^{\rm
h}54^{\rm m}27^{\rm s}\!.51$, $+41^{\circ}34'46''\!.29$), is slightly
shifted to southeast from the 2MASS image center of NGC 1129, (R.A.,
decl.)$_{\rm J2000}$ = ($2^{\rm h}54^{\rm m}27^{\rm s}\!.34$,
$+41^{\circ}34'46''\!.20$), by $1.9''$ (1 kpc) with a statistical
error of $0.3''$ as shown in Figure 2c. This feature has been noted by
\cite{kg00} who reported that the peak of the PSPC image was $6''$
offset from the nominal position of NGC~1129. In the 2--10 keV band,
coordinate of the X-ray peak at the cD galaxy is (R.A., decl.)$_{\rm
J2000}$ = ($2^{\rm h}54^{\rm m}27^{\rm s}\!.40$,
$+41^{\circ}34'46''\!.93$), which is $1.0''$ (0.5 kpc) away from the
2MASS image center. However, the statistical error in this band is 
$1.5''$ so the uncertaintly range covers both the 2MASS and the
soft peaks. Figure 2d shows one-dimensional profiles from southeast to
northwest running through the hard sub-peak, and the center of the cD
galaxy. The peak of the soft-band profile (open circles) is offset by
a few arcsec, but the hard band one (filled circles) takes the maximum
almost at the cD center.

\subsection{Fe Blobs}
\label{sec:feblob}

% Figure 3 : Fe band image and hardness
The left panel of Figure 3 shows an image by limiting the data in a
narrow energy band 6--7 keV, which includes the Fe-K line, for the
central $1'$ region. The image shows two blob-like features
symmetrically located around the cD galaxy center in the northeast and
southwest directions. The two blobs contain 25 and 33 counts in the 6--7 keV 
band, while the surrounding regions of the same size have 9 and 13 counts, 
respectively. Interestingly, the angle of the line connecting
the two blobs is perpendicular to that connecting the cD galaxy center
and the hard sub-peak as seen in the 2--10 keV intensity contours. To
confirm the enhanced Fe abundance, we calculated hardness ratios
between fluxes in 6--7 keV and 2--6 keV for all positions and plotted
them in the right panel of Figure 3, by converting them into the
abundance values. Since the ICM temperature at $r=10''-20''$ is fairly
constant within 20 \% in this region, the hardness ratio should show
us the approximate equivalent width of Fe-K line excluding the very
center. We took running means over $16'' \times 16''$ for each data
point, because the photon statistics in the 6--7 keV band was too low
to draw a pixel-by-pixel image. The region with good statistics is
encircled by the white line, in which the statistical errors are less
than 30\%.  The hardness map also indicates high-abundance regions
around the two blobs seen in the left panel.  As shown in the spectral
fit in \S\ref{sec:2dmap}, the abundance value of the Fe blobs is
1.5--2.0 solar.  We also note that in both panels of Figure 3 the
southeast hard sub-peak has a very low Fe abundance, although the
statistical errors are large.

\subsection{Spectral Analysis}
\label{sec:cDspec}

We first extracted pulse-height spectra for the central
three regions, which correspond to the hard peak (cD center), soft
peak, and the hard sub-peak as shown in Figure 2b. The region sizes are $7''\times 14''$,
$7''\times14''$, and $10''\times10''$, respectively. The data were
corrected for CTI using the software provided by Townsley
\footnote{http://www.astro.psu.edu/users/townsley/cti/install.html},
which significantly improved residuals of the spectra around Fe-K line.
We have created background spectra using the blank sky data described by
Maxim Markevitch \footnote{http://asc.harvard.edu/cal}.  Response
matrices and effective area files were made for multiples of
$32\times 32$ pixels, and averaged with a weight of the number of photons
in the region  using ``mkrmf'' and ``mkarf'' in the CIAO software
package distributed by the CXC\@. Since the response for the ACIS spectrum
has a problem below 1 keV in determining the accurate
absorption due to the material accumulated on the ACIS optical blocking
filter, we added the {\sc acisabs} absorption model
\footnote{http://legacy.gsfc.nasa.gov/docs/xanadu/xspec/models/acisabs.html}
available in XSPEC, and fixed the absorption at the Galactic value of
$9\times10^{20}$ cm$^{-2}$. The parameter {\sc Tdays} was set to 392,
and the rest of the parameters were kept at their default values.  We
fit the spectra with an absorbed single temperature MEKAL model in the
energy band 1--9 keV, excluding the 1.8--2.2 keV interval around the 
mirror Ir edge to minimize the effects of calibration uncertainties.

% Figure 4 : Spectra of the central region
% Table 1 : The parameters

The spectra and resulting best-fit parameters are shown in Figure 4
and Table 1. The normalizations of the spectra are arbitrarily scaled
to lay-out the spectra separately. The spectra of the soft and hard
peaks, where most part of the emission comes from the NGC 1129 center,
show the temperatures of 1.8 and 2.0 keV.  They are consistent with
the previous PSPC value of $kT=1.8$ keV at the cluster center
\citep{nb95} and are about half of the ICM temperature in the outer
region.  The luminosity of the cD galaxy, by adding the former two
regions, is derived to be $2.6\times 10^{41}$ erg s$^{-1}$ in the
0.5--10 keV band. The temperature of the hard sub-peak is about 3 keV,
higher than the former two regions just as expected from the color
X-ray image (Figure 1b).  Since the spectrum of the sub-peak contains
the hot ICM component along the line of sight, a spectral fit with an
additional MEKAL component with a fixed temperature of 2.8 keV, which is 
roughly an average temperature at $r=10''-20''$ as shown in Figure 6, was 
carried out and resulted in the sub-peak temperature to be higher than
about 5 keV with the best-fit metallicity value of zero solar. The
metal abundance of the sub-peak is significantly lower compared with the
other 2 regions and the surrounding ICM (see
\S\ref{sec:ktz}). Accordingly, the data also allows a power-law
spectrum for the sub-peak component, with the best-fit photon index
1.2 and the 90\% error range to be 0--2.5. However, the statistical
quality of the data is poor and it is difficult to constrain the
actual nature of the sub-peak emission.  The estimated luminosity of
the sub-peak excluding the ICM component is $1.2\times 10^{40}$ erg
s$^{-1}$, which is close to the levels of small elliptical galaxies.

To evaluate the significance of the enhanced Fe abundance, we also
created a pulse-height spectrum for the sum of the two Fe blob regions
and compared it with the one in the surrounding region. The actual
regions where the data were accumulated are shown by the white
ellipses in Figure 3a for the Fe blobs, and the surrounding low-metal
region was an annular region with $r=5''-20''$ excluding the Fe
blobs. Figure 5 shows the resultant spectra for the Fe blobs and the
surrounding region. A prominent Fe-K line is indeed recognized in the
blob spectrum. We fit the spectra in the same way as we did for the
central spectra, i.e.\ single temperature MEKAL model with a fixed
$N_{\rm H}$ value.  The Fe blob region shows the metal abundance to be
1.57 (1.03--2.10) solar, compared with 0.73 (0.52--1.04) solar for the
surround region, with the values in parentheses showing 90\%
confidence limits for a single parameter. We looked into the
difference in the $\chi^2$ values for two spectral fits, one with a
common abundance for the two spectra and the other with separate
abundance parameters, and it turned out to be 8.1 (138.8 and
146.9). The reduction of the $\chi^2$ value is more than 2.7 for one
additional model parameter, which indicates that the two regions have
different metal abundance at the 90\% confidence \citep{malina76}.

\section{THE SURROUNDING ICM REGION}

\subsection{Surface Brightness Profile}
\label{sec:SB}

% Figure 6 (top) : Surface Brightness
The top panel of Figure 6 shows the radial distribution of the X-ray
surface brightness within a radius of $r<2'$(60 kpc) in the 0.5--10
keV band. Point source candidates were subtracted, and the exposure
was corrected for. We center on the X-ray peak of the 0.5--10 keV
image, which is the same position as the soft-band peak because the
photons below 2 keV are dominant even in the full band image.  The
X-ray emission from the cD galaxy within a radius of $\sim20''$ is
smoothly connected to the ICM without significant jump or edge
features. The profile as a whole is very well fitted by a single
$\beta$ model with the parameters derived as $\beta=0.28 \pm 0.01$ and
$r_{\rm c}=1.4''\pm 0.2$ (0.7 $\pm$ 0.1 kpc). Errors denote the 90\%
confidence limits for a single parameter. The $\beta$ value is
slightly larger than the previous HRI results which showed $\beta=0.25
\pm 0.01$, and the core radius is smaller than the HRI value, $r_{\rm
c}=6''-16''$ \citep{nb95}. The difference is reasonably explained by
the higher spatial resolution of the {\it Chandra} data.  The $\beta$
value obtained here is much lower than those in typical cD galaxies
($\beta=1-2$) or field elliptical galaxies ($\beta=0.4-0.6$). The
central density is obtained to be $n_0=0.053\pm 0.010$ cm$^{-3}$,
which is consistent with the value derived from the HRI observation.

\subsection{Temperature and Abundance Profiles}
\label{sec:ktz}

% Figure 6 (lower 2 panels) : kT and Z profiles
We have created annular spectra with a width of $10''$ and fitted them
with an absorbed MEKAL model. The point source candidates are
subtracted. We have created background spectra in the same way as we
performed for the spectra of the central region in \S\ref{sec:cDspec}.
Each annular spectrum can be fitted well by an absorbed single
temperature model with a reduced $\chi^2$ of mostly 0.9--1.2. The
middle panel of Figure 6 shows the resulting temperature profile. We
fit the spectra for two different energy bands, 1--9 keV and 2--9 keV,
as indicated by crosses and diamonds in the lower two panels of Figure
6. The best-fit temperatures for these two energy bands are slightly
different in each ring by about 0.5 keV\@. The tendency that the temperature gradually drops toward
the center from around 4 keV is the same for both fits.  The central
temperature is 2 keV, about half of the average ICM temperature in the
outer region. Similar features are seen in most cD clusters with ASCA
(Ikebe 2001), and is also similar for a universal temperature profile
for relaxed lensing clusters reported by \cite{allen01}.

The bottom panel of Figure 6 shows the radial profile of metal
abundance.  The abundance increases toward the center from 0.5 solar
to 1.5 solar at $r=7$ kpc. The peak value exceeding 1 solar is much
higher than the previous ASCA result \citep{ezawa97} which is
indicated by the dotted line.  The abundance feature was not resolved
by the large point spread function of the ASCA X-ray mirror. The
central abundance falls back to 0.7 solar with the single temperature
fit in the 1--9 keV band, leaving the ring-like metallicity peak at
$r=7$ kpc. The similar profile is observed in other cD clusters, such
as in the Centaurus cluster \citep{sf02}, Abell 2199 \citep{johns02},
and Abell 2052 \citep{blant03}. To take the projection effect into
account, we also fitted the central spectrum with a two-temperature
model with fixed parameters of the projected ICM component as shown
with the thick crosses. The resulting best-fit abundance is about 1.2
solar, which is still lower than the peak value. However, the errors
now allow the curve to be flat, or a metal concentration at the center
is allowed if one takes the result (1.6 solar) for the 2--9 keV
fit. The deprojected abundance profile of M~87 shows the same result,
which decreases to the center in the whole band fit, but increases to
the very center in the fit above 2 keV \citep{matsu02}. Unfortunately,
the smaller core of AWM~7 with rather low surface brightness, compared
with the one of Centaurus, M~87, etc, makes confirmation of the low
abundance at the very center of AWM~7 difficult.

\subsection{Two-dimensional Map}
\label{sec:2dmap}

% Figure 7
Since the radial profile always suppresses fine angular structures, we
studied the spectra in 4 azimuthal directions separately for the
central $2'\times2'$ region in steps of $\Delta r=10''$.  We fit the spectra with a single
temperature model in the 1--9 keV band. Figure 7 shows the resulting
2-dimensional maps of temperature and abundance. The map shows that
the temperature in the west half is almost isothermal with $kT = 3.0-3.5$
keV excluding the central region with $r < 20''$. The southeast region
is hotter than the average by 1.0 keV, while the surface brightness
distribution is very smooth and symmetric.  The abundance map again
shows the high metallicity regions in the northeast of the second ring
and the southwest of the third ring. These two regions exactly
correspond to the two blobs described in \S\ref{sec:feblob}.  The
best-fit abundances of the northeast and southwest blobs are $1.9\pm
0.5$ solar and $2.8^{+0.5}_{-0.9}$ solar, respectively, while the
surrounding regions show 1.0--1.4 solar with typical errors of
0.2--0.4 solar. The abundance distribution is almost uniform with
$Z=0.7-0.9$ solar in the four directions outside the Fe blobs.

\section{DISCUSSION}

We have analyzed the {\it Chandra} data of the galaxy cluster AWM~7, which
brought us the X-ray view of the cluster center with the best spatial
resolution ever achieved. In particular, the 2--10 keV image showed
interesting structures in the cD galaxy. We found an extended emission
with fairly hard spectrum ($kT \gtrsim 3$ keV) at a position 6--7 kpc
off from the cD center. This source has no optical counterpart. Other
weaker blob-like features are also seen in the 2--10 keV image in the
central region.  Based on the Fe band intensity map and the hardness
ratio map, we found two extended regions with strong Fe-K line
emission almost symmetrically aligned with respect to the cD
center. The directions of these Fe blobs from the cD center are
perpendicular to that of the hard sub-peak.  We derived radial
profiles of the surface brightness, temperature, and metal abundance
for the surrounding ICM within a radius of $2'$, which showed some
common X-ray features with other cD clusters. Below, we will discuss
about the implications of these new and interesting results.

\subsection{Hard Sub-Peak and Hot Blobs}

The sub-peak recognized in the hard band image has a spatial extent of
about 4 kpc if it belongs to AWM~7, and it is certainly more extended
than the PSF\@. The X-ray luminosity at the distance of AWM~7 is
estimated to be $1.2 \times 10^{40}$ erg s$^{-1}$, which is comparable
to those of elliptical galaxies. Other fainter blobs which are
suggested in the hard band image (Figure 2b) have lower luminosities
by several factors. Here, we examine the possibility whether this hard
sub-peak really belongs to AWM~7 or not. The observed properties of
this source, i.e.\ an extended X-ray emission with a temperature
$\gtrsim 3$ keV (or nonthermal) and the lack of optical counterpart,
makes the possibility of foreground or background object very
unlikely.  Possible extragalactic objects with extended
emission are elliptical or starburst galaxies, or distant clusters,
but the absence of optical counterpart is a severe
problem. Furthermore, the complex structure seen in the hard-band
image strongly suggest that the whole feature is the emission of AWM~7
itself and that the hard sub-peak is simply the brightest blob among
the other fainter ones. With these consideration, we may conclude that
the sub-peak feature is a gas cloud residing in the core region of
AWM~7\@. We note that a number of X-ray blobs with a similar size (3
kpc) have been found in the core region of the cluster 2A~0335+096
\citep{mazz03}, in which blob luminosities are several times $10^{41}$
erg s$^{-1}$. A notable difference is that the blob temperatures in
2A~0335+096 ($\sim 1.5$ keV) are similar to the surrounding ICM level,
while in AWM~7 the blobs seem to be hotter or the emission is
nonthermal.

Let us look into the nature of the hot blobs.  If the emission is
thermal, the temperature of the sub-peak is higher than 3 keV and the
metal abundance is relatively low at $\lesssim 0.3$ solar. These
properties are quite unique, and do not match the gas properties of
either elliptical galaxies or surrounding ICM\@. Since there is no
radio emission observed at the cluster core, non-thermal pressure such
as due to magnetic field is unlikely to be a major source to confine
the hot gas in the sub-peak component. In that case, the hard sub-peak
and other fainter blobs may not be in pressure equilibrium with the
surrounding gas. The sound crossing time to smooth out the density
structure is $t_{\rm cross}=9\times 10^6$ yr, which is shorter than
either the conduction time of $1.3\times10^7$ yr, or the cooling time
of $t_{cool}=7\times10^8$ yr. If there is indeed no external pressure
which stops the structure dissipating away, these hot blobs have to be
created very recently, within 1--10 Myr. If major part of the hot gas
in the central region is moving, as suggested from the sharp-edge
feature in Figure 2b, the blobs may be continuously created in the
turbulence.

\subsection{Metal Distribution}

The projected radial profile of the metal abundance shows a steep
gradient, reaching about 1.5 solar. The abundance may fall back at the
center, and apparently creates a high metallicity ring at $r=10''-20''$
(5--10 kpc). The off-center peak in the abundance profile was found in
some major cD cluster cores.  The radius of the metallicity peak is,
however, smaller in AWM~7 than those in other clusters, such as $r=15$
kpc for Centaurus \citep{sf02}, $r=30$ kpc for Abell 2199
\citep{johns02}, and $r=60$ kpc for Perseus \citep{sfs02}. \citet{mf03}
suggests that the central abundance begins to decrease with the thermal
evolution of ICM with a non-uniform metal distribution, which can
account for the smaller peak radius of the rather young cluster AWM~7.

Our 2-dimensional hardness analysis showed that the observed high
metallicity peak was mainly caused by the two Fe blobs symmetrically
located around the cD center. The angles of the Fe blobs from the cD
center are perpendicular to hard sub-peak direction.  This
configuration suggests that the high metallicity gas may have been
expelled from the cD galaxy due to some pressure invoked along the
northwest to southeast direction. The hard sub-peak and the fainter
one on the other side may be causing such a pressure, for example, by
falling into the cD galaxy on both sides. If this is the case, the
relatively higher temperature of the blobs can be explained as due to
shock or compressional heating.  

The total mass of Fe contained in the Fe blobs is estimated to be
$2\times10^4\ M_{\odot}$. The star formation rate of NGC~1129 is $0.04
M_{\odot}$ yr$^{-1}$ within $r=1.6$ kpc \citep{mcoc89}.  Assuming that
1 solar mass of Fe is injected into ICM when a total of 100 solar mass
of the stars are formed, then $M_{Fe}=4\times 10^5\ M_{\odot}$ has
been produced within 1 Gyr. Only 5\% of the total Fe
mass is enough to account for the Fe blob. Therefore, at least the
amount of Fe in the Fe blobs can be explained by the metal production
in the cD galaxy, even though the reason why metal injection occurs in
a directional way is not clear.

\subsection{Gas Heating in the cD Galaxy}

The radial temperature distribution shows a smooth and monotonic
decrease from 4 keV to 2 keV at the cluster center. The temperature
decrease toward the cluster core has been previously recognized as
cooling flows, but the data from {\it XMM-Newton} showed that the
low-temperature gas with $kT \lesssim 1$ keV is missing in many cD
clusters. In AWM~7, the radiative cooling time at the center is
estimated to be $t_{cool}=3\times 10^8$ yr, which is shorter than the
cluster age. However, the central temperature seems to be maintained
at 2 keV, which is about half of the gas temperature in the outer
region. This situation is similar to the feature seen in many other cD
cluster centers observed with {\it Chandra} and {\it XMM-Newton}
\citep[e.g.][]{peter02}.  It seems that in AWM~7 also, the thermal
structure in the cluster center is determined by a certain balance
between radiative cooling and some heat source.  We note that the same
problem about the unknown heat source so far recognized in cooling flow
systems is also met in this relatively poor cluster. 

Several scenarios to account for the heat source in cluster cores are
proposed by several authors.  One of the heating scenario is to invoke
radio jets from the central AGNs \citep{hans02}. In the case of AWM~7,
however, no radio activity has been observed in NGC 1129, which makes
the AGN jets unlikely to be the heat source. Another possibility to
suppress the cooling is the heat input through thermal conduction from
the surrounding hotter ICM gas. The necessity for unrealistic fine
tuning of the heat conductivity over a wide region is found in
explaining the observed temperature distribution \citep{kita03}.
Therefore, heat conduction is difficult as the
heating mechanism.

In the present case, we consider that the gas motion in the central
region may be at least related to the heat source. As already shown
earlier, there are hot blobs which are hotter than the gas in the cD
center. We also discussed that these blobs may have been created by a
possible bulk motion of the gas in the central region, as suggested by
the morphology in the hard energy band. Although we do not know the
detailed feature and the exact cause of such a motion, it is enough to
heat up a considerable fraction of the gas up to $\gtrsim 3$ keV as seen
in the sub-peak spectrum. If this energy is brought in to the central
cool gas, it would certainly be able to compensate the energy carried
away by the radiative cooling. 

The position of the soft X-ray peak shows a significant offset by $1
\pm 0.2$ kpc from the 2MASS position of the cD galaxy center, whereas
the X-ray peak in the hard band suggests a smaller offset by $0.5 \pm
0.8$ kpc.  The offset between the soft X-ray and the 2MASS peaks
suggests that there is a relative motion between the hot gas and the
galaxy core.  In the optical band, there is a chain of four small
galaxies extending in the southwest direction, so the galaxy
distribution is out of symmetry around the cluster core.
\citet{pele90} reported that there is a large twist in the position
angle of NGC 1129 at a radius of $10''-20''$ in the optical band. They
suggest that this twist could be a result of a galaxy merger. With
these optical features as well as the offset of the soft X-ray peak,
we may consider that the central region of AWM~7 is fairly dynamic and
that it may well connected with the unidentified heat source.

Since ASCA showed a Mpc-scale abundance gradient and isothermal gas
distribution, it has been thought that AWM~7 has not experienced
cluster-scale mergers nor significant mixing. The present {\it
Chandra} data revealed kpc-scale structures and suggested that the
central region can be dynamic. Further direct information about the
dynamical motion of the gas blobs (hard sub-peak and Fe blobs) will
give us clearer view about the actual physical process occurring in
the cluster core. X-ray imaging spectroscopy facilitated by a new
technique, such as microcalorimeters on Astro-E2, will bring us rich
information about the evolution of cD galaxies and clusters.

\acknowledgments

We would like to thank Dr.\ Y. Ikebe, and the referee for useful
comments. T. F. is supported by the Japan Society for the Promotion
of Science (JSPS) Postdoctoral Fellowships for Research Abroad. This
work was partly supported by the Grants-in Aid for Scientific Research
No.\ 12304009  from JSPS\@. NASA/IPAC Extragalactic
Database (NED) is operated by the Jet Propulsion Laboratory,
California Institute of Technology, under contract with NASA and the
SIMBAD data base is operated by the Centre de Donnes astronomiques de
Strasbourg.

%%%%%% Table 1 %%%%%%%%%%
\begin{table}
\begin{center}
\caption{Best-fit parameters of the spectral fit for the 3 selected
regions}
\begin{tabular}{lccccc} \hline \hline 
region        & offset$^\dagger$ & $kT$   & $Z$  & $\chi^2$/dof  & $L_{\rm X}$ $^\ddagger$ \\
              & (kpc) & (keV)   & (solar)  & &  (erg/s)  \\
\hline
soft peak     &  2.0 & 2.02$^{+0.12}_{-0.14}$ & 0.89$^{+0.28}_{-0.24}$ & 100/81 & $1.2\times 10^{41}$ \\ 
hard peak (cD center)  &  0.8 & 1.79$^{+0.19}_{-0.06}$ & 0.68$^{+0.16}_{-0.14}$ & 76/81  & $1.4\times 10^{41}$ \\
hard sub-peak &  6.4 & 2.94$^{+0.35}_{-0.19}$ & 0.34$^{+0.18}_{-0.15}$ & 104/77 & $1.2\times 10^{40}$ \\
\hline
\end{tabular}
\end{center}

\noindent $^\dagger$ Distance from the center of the cD galaxy.

\noindent $^\ddagger$ X-ray luminosity in the 0.5--10
keV band assuming that the distance is 104 Mpc.
\end{table}

%%%%%% Figure 1 %%%%%%%%%%

\begin{figure}[htb]
\begin{center}
\plottwo{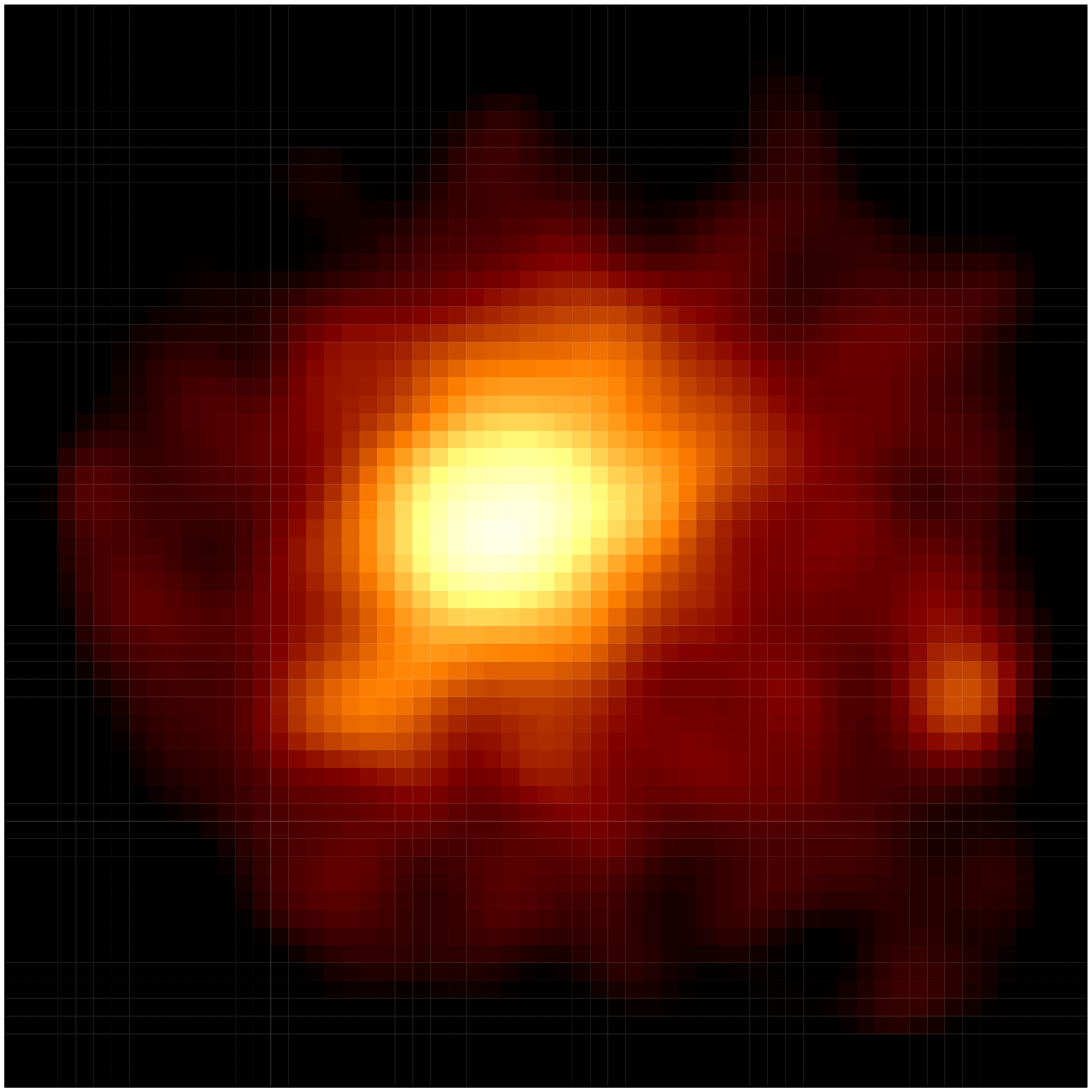}{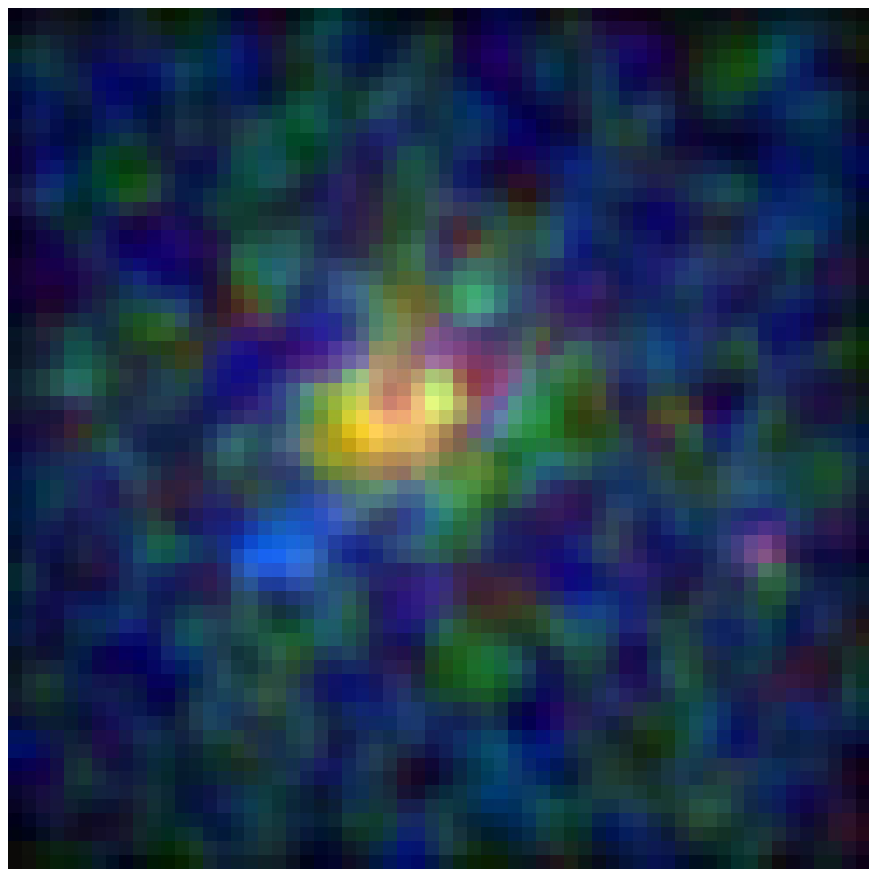}
\caption{(a) {\it Chandra} image of the central $1'\times1'$ region of
     AWM~7 in the 0.5--7 keV band. The image was smoothed by a Gaussian function with $\sigma = 2''$. (b) X-ray color image composed of 0.5--1.5 keV (red),
     1.5--2.5 keV (green), and 2.5--8 keV (blue) images. The pixel size is 1 arcsec.}
\label{0.5-7keVimage}
\end{center}
\end{figure}

%%%%%% Figure 2 %%%%%%%%%%

\begin{figure}[htb]
%%\vspace*{-.5cm}
\hspace*{1cm}(a)\hspace*{7.5cm}(b)\hspace*{7.5cm}
\vspace*{-4mm}
\begin{center}
\plottwo{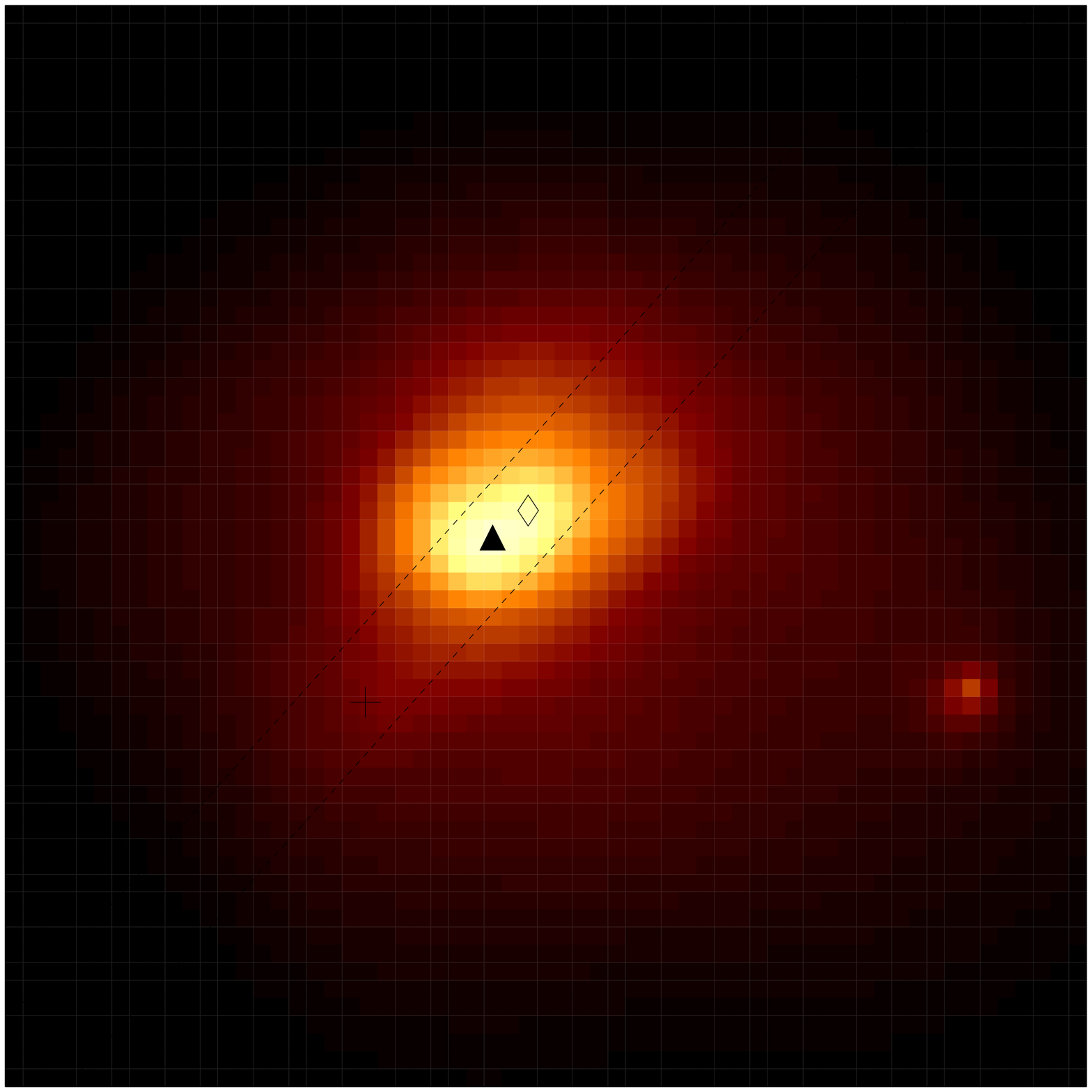}{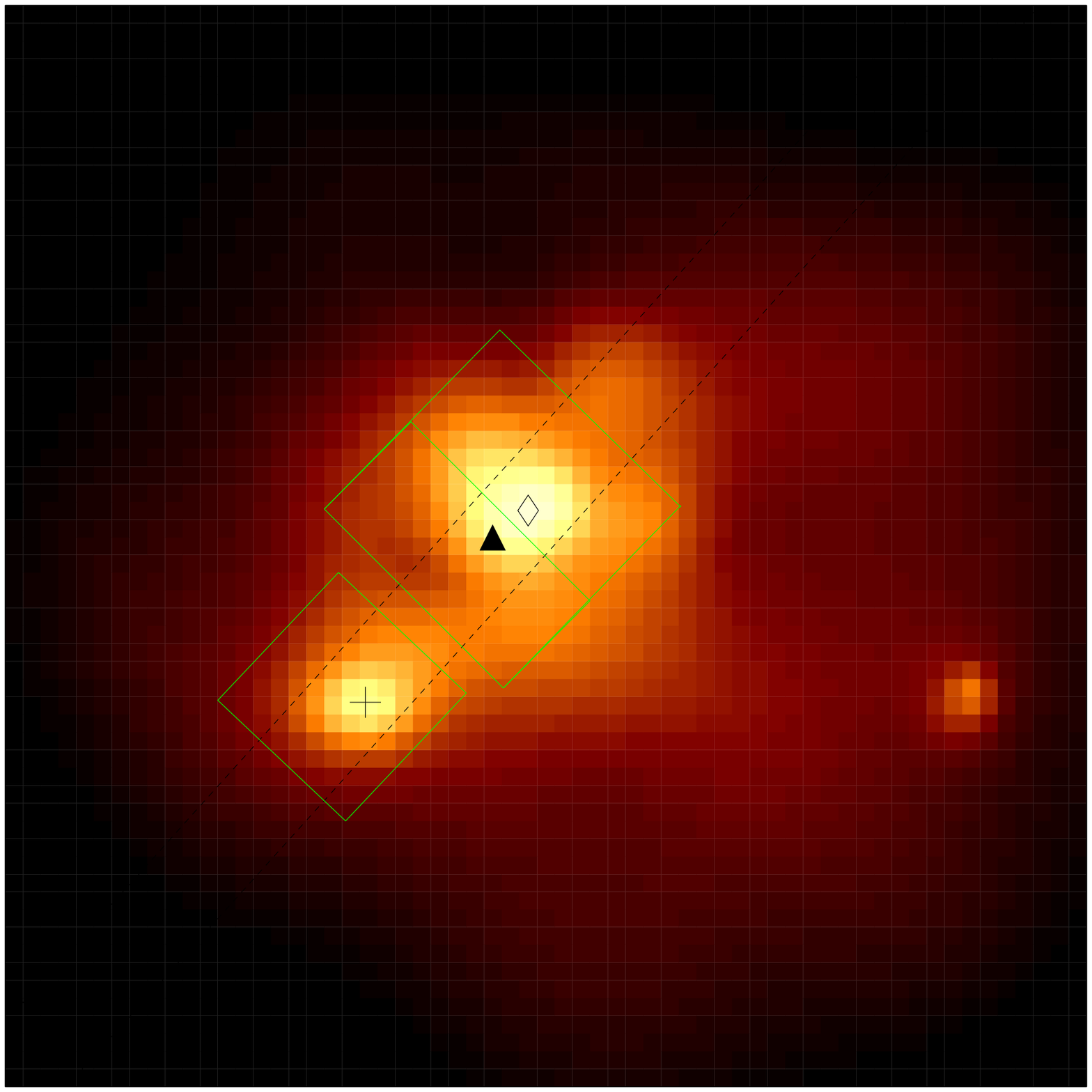}
\end{center}
\end{figure}
%%\vspace*{-7cm}
\begin{figure}[ht]
\hspace*{1cm}(c)\hspace*{7.5cm}(d)\hspace*{7.5cm}\\
\vspace*{-8mm}
\begin{center}
\plottwo{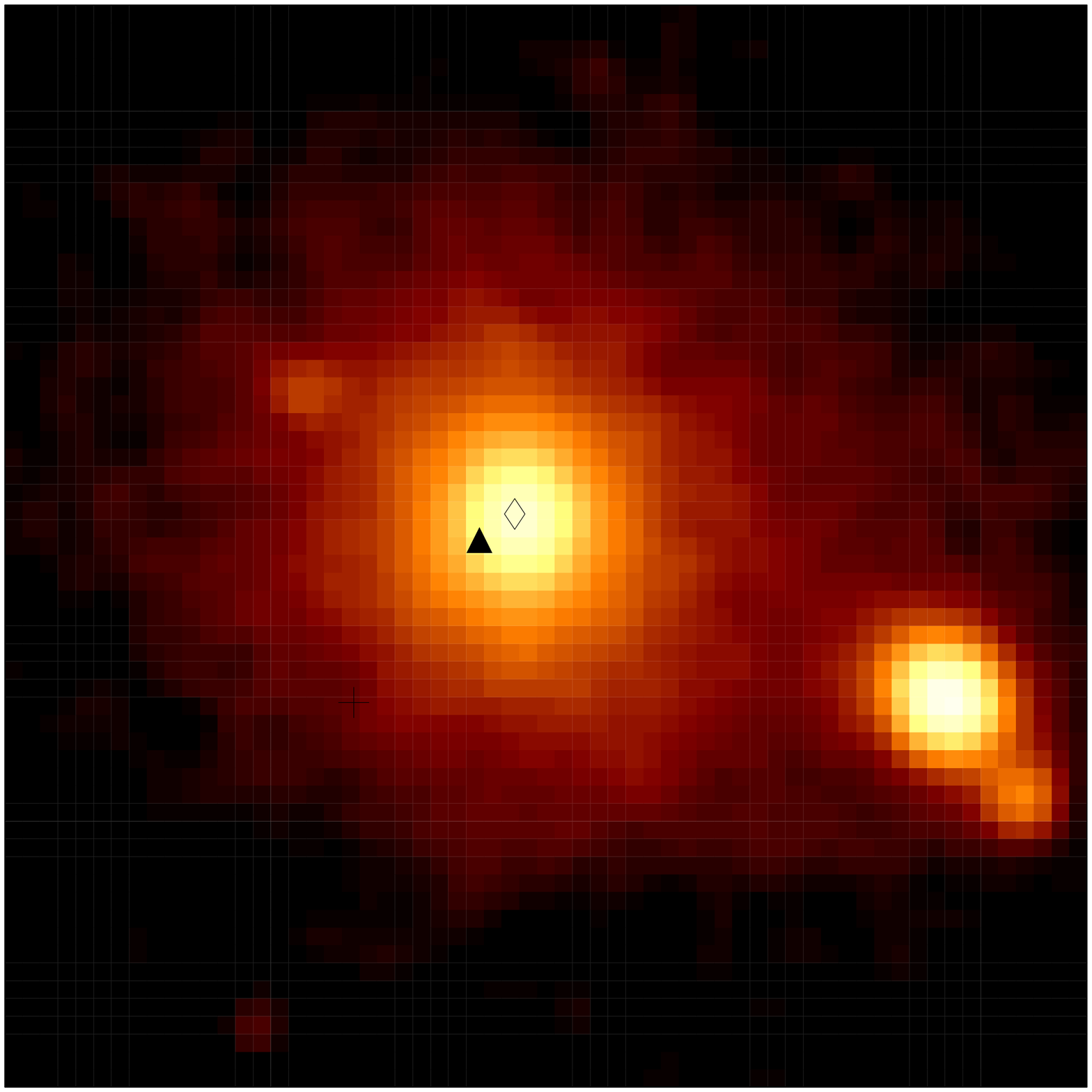}{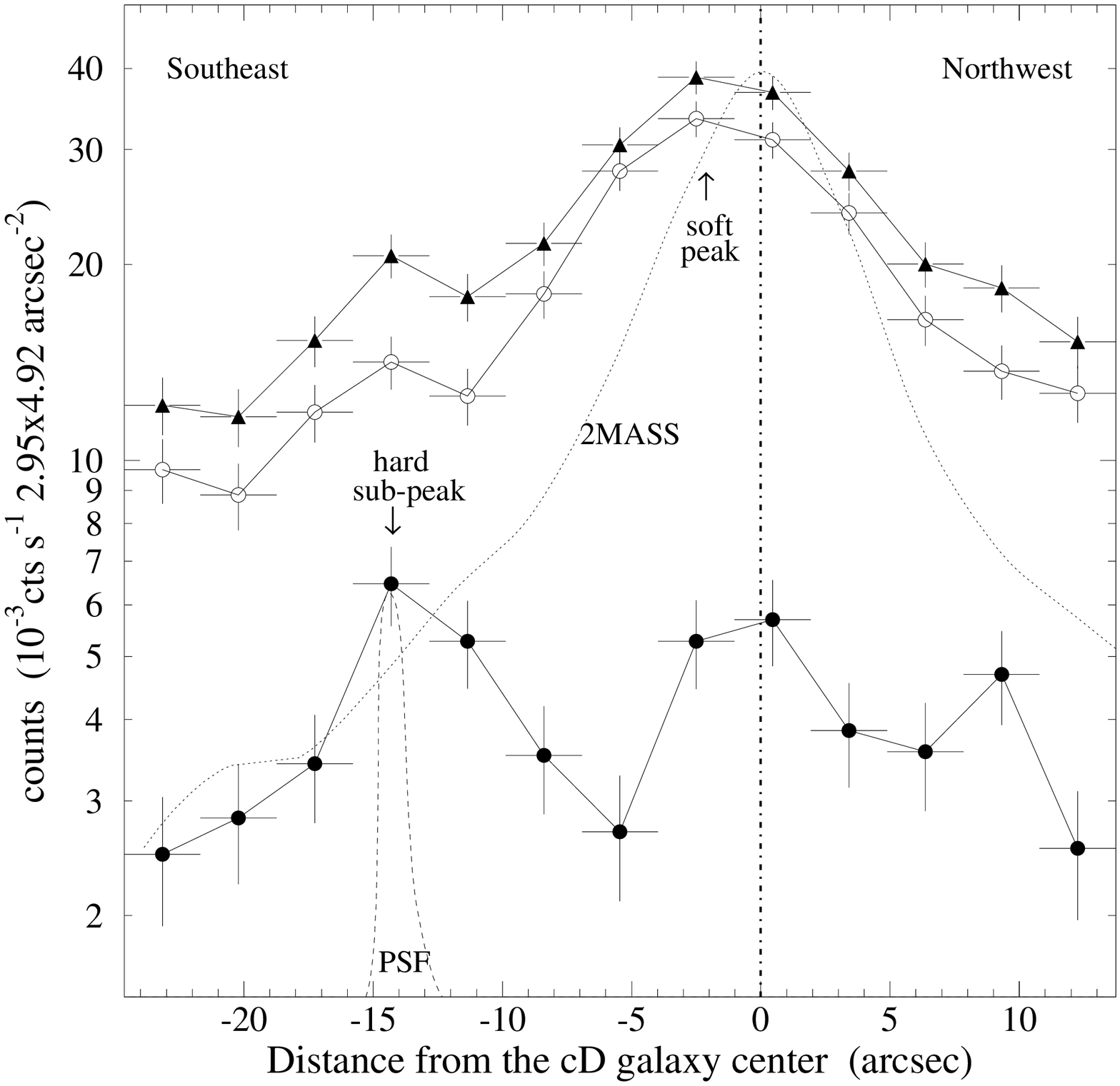}
\caption{ (a) Chandra images in the 0.5--2 keV, (b) 2--10 keV, and (c)
the 2MASS image.  The images are 1 arcmin square with a pixel size of 1 arcsec. The triangle indicates
the peak position in the 0.5--2 keV band, and the diamond and cross
indicate the position of the cD galaxy and hard sub-peak. The boxes in (b) show the regions where the spectra in Figure 4 were extracted. (d) Image
profiles from southeast to northwest across the hard sub-peak, and the
cD galaxy center as a function of distance from the cD galaxy
center. The curves are of the 0.5--10 keV, 0.5--2 keV, and 2--10 keV
bands from upper to lower with a projected width of $5''$. The dotted
and dashed histograms represent normalized profiles of the 2MASS image
and PSF at the hard sub-peak. }
\label{images}
\end{center}
\end{figure}

%%%%%% Figure 3 %%%%%%%%%%

\begin{figure}[htb]
\begin{center}
\plottwo{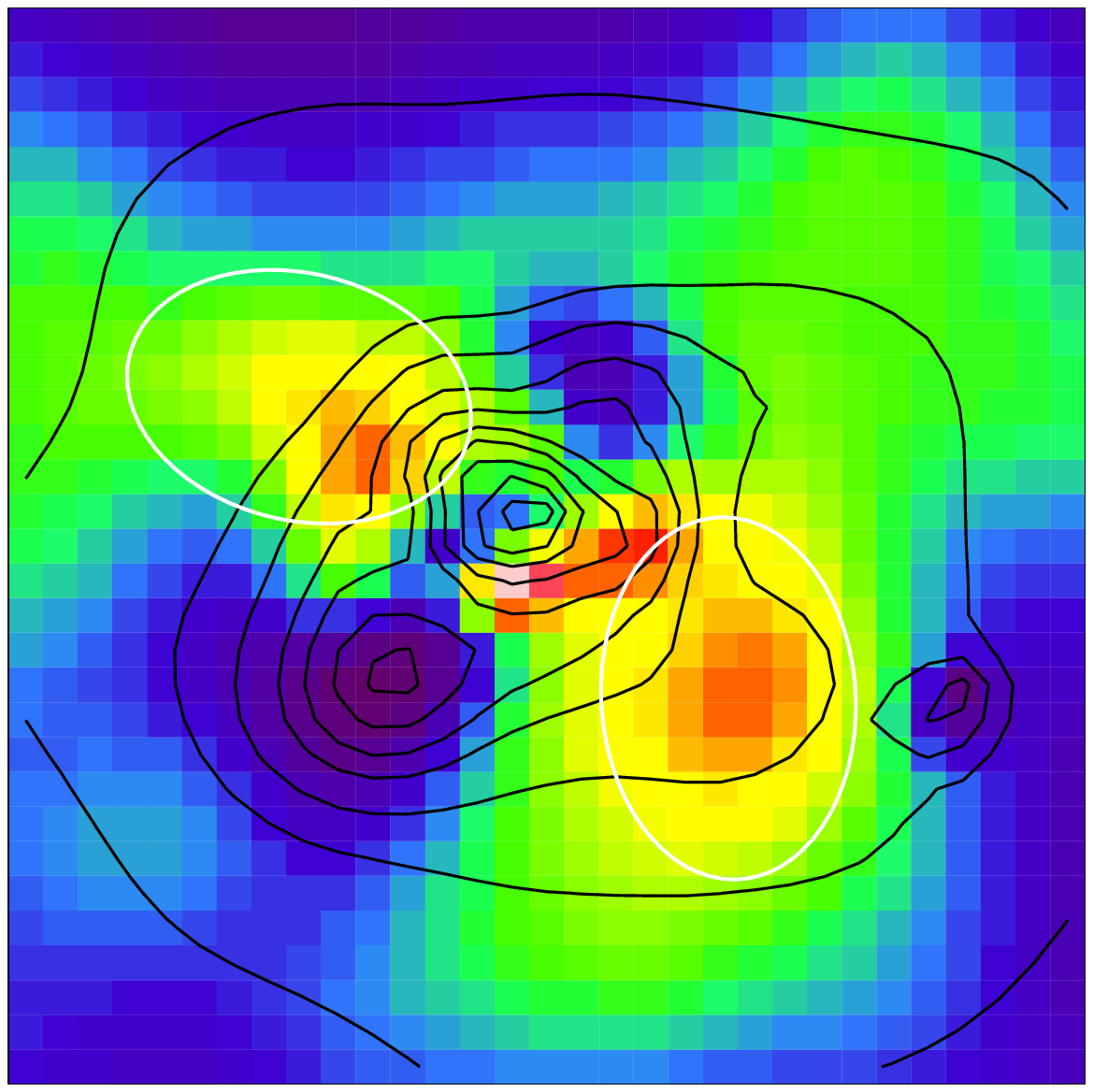}{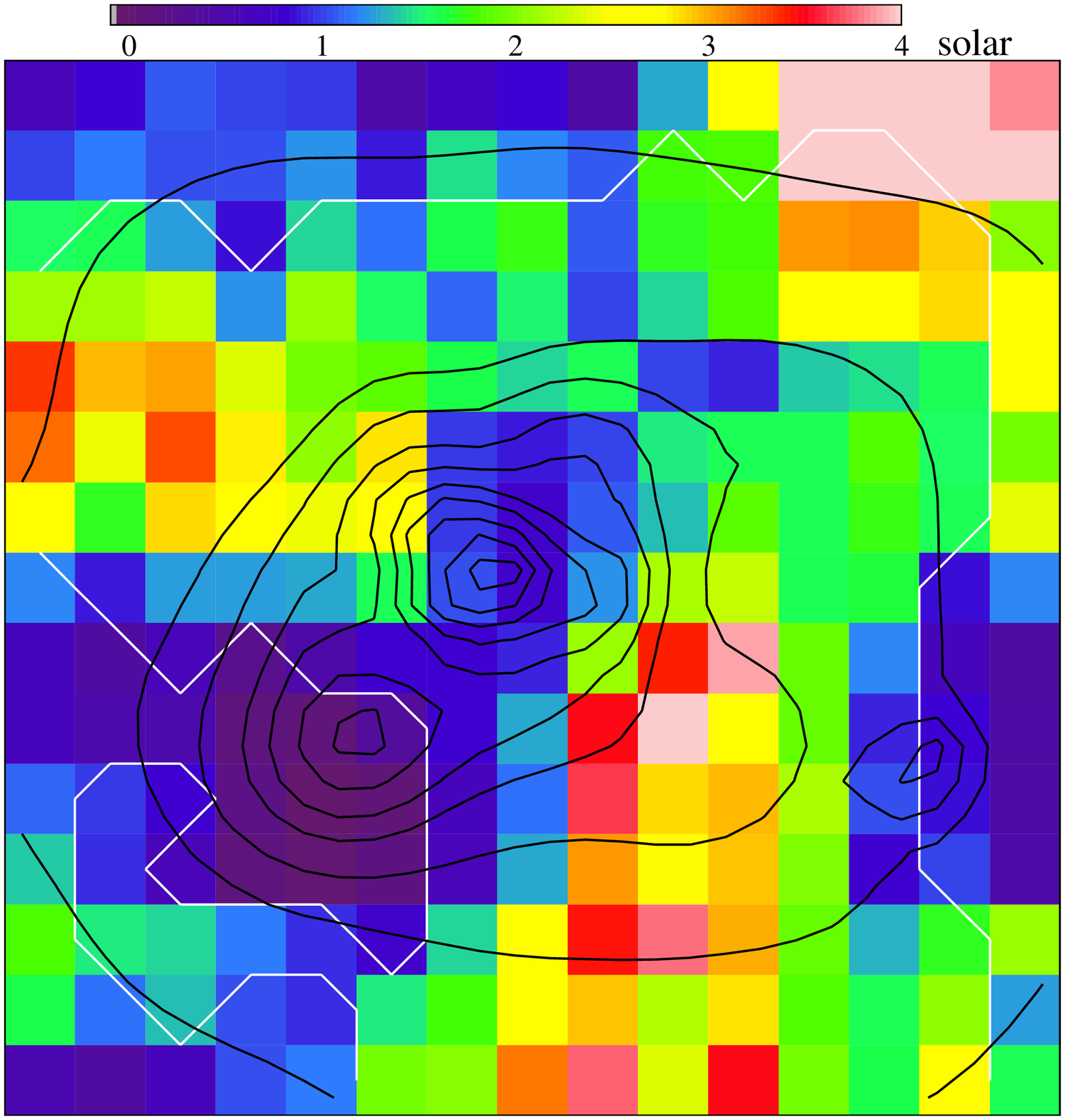}
\caption{{\it Left)} The 6--7 keV image of the central $1'$ square
region with the 2--10 keV intensity contours. The white ellipses
indicate the Fe blob regions to extract a spectrum shown in Figure
5. The pixel size is 2 arcsec. {\it Right)} Fe-K band hardness map of ratios of the 6--7 keV image
to the 2--6 keV image of the central $1'$ region (see text). The red/blue tone
indicates more/less Fe-K band counts. Inside of the white line
represents where the statistical errors are less than 30\%. The pixel size is 4 arcsec. The contours
indicate the hard (2--10 keV) X-ray image. }
\label{fekimg}
\end{center}
\end{figure}

%%%%%% Figure 4 %%%%%%%%%%

\begin{figure}[htb]
\begin{center}
\plotone{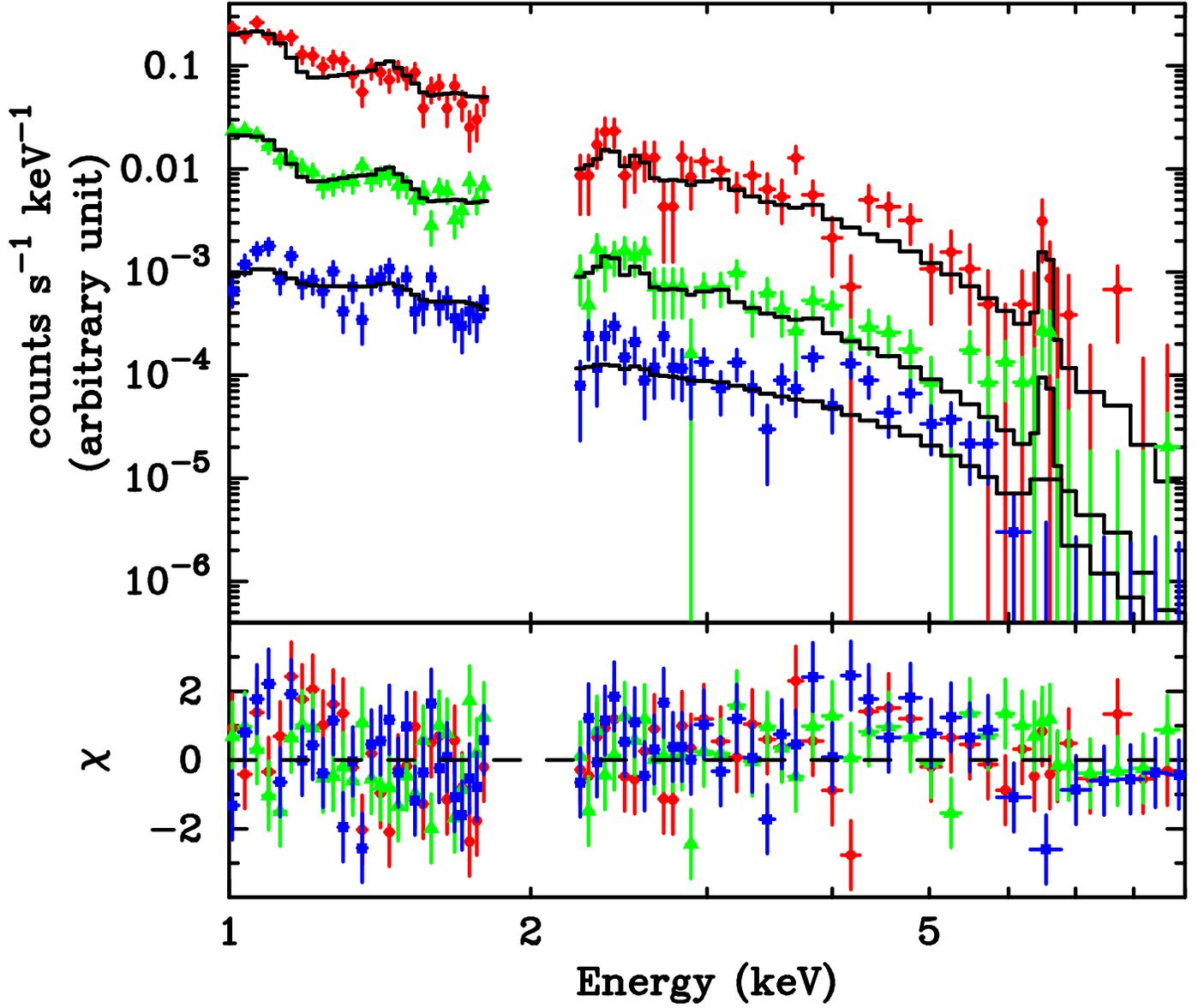}
\caption{Spectra of the very central region of AWM~7. From the top, the
hard and soft peak regions of the cD galaxy center, and hard sub-peak
region as shown in Figure 2b (see also text). }
\label{spectra}
\end{center}
\end{figure}

%%%%%% Figure 5 %%%%%%%%%%

\begin{figure}[htb]
\begin{center}
\plotone{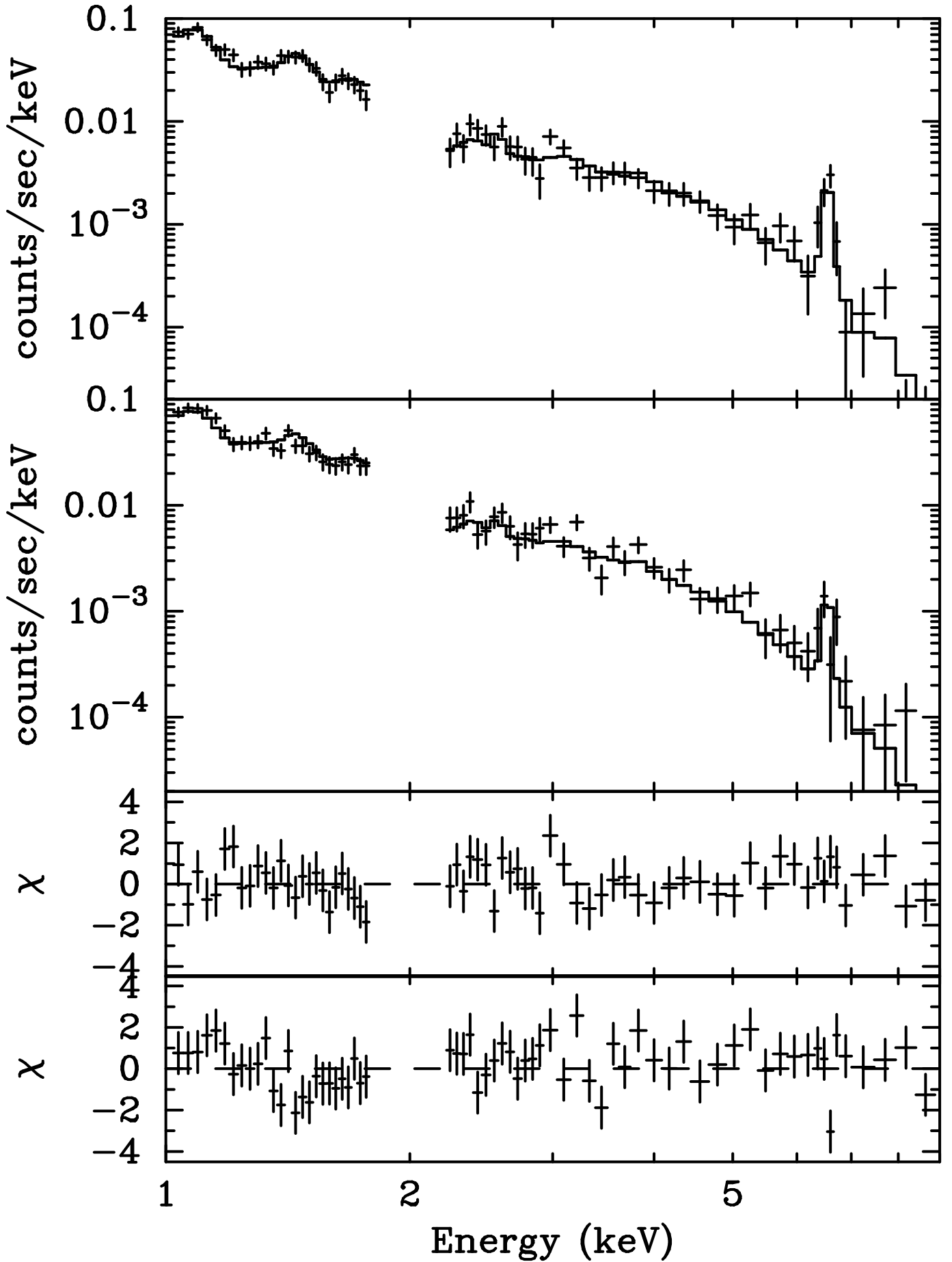}
\caption{Spectra of the Fe blob region as shown in Figure 3a (top) and
the surrounding $r=5''-20''$ region (middle), fitted with a model of
Galactic absorption and a MEKAL component. }
\label{spectra}
\end{center}
\end{figure}

%%%%%% Figure 6 %%%%%%%%%%

\begin{figure}[htb]
\begin{center}
\plotone{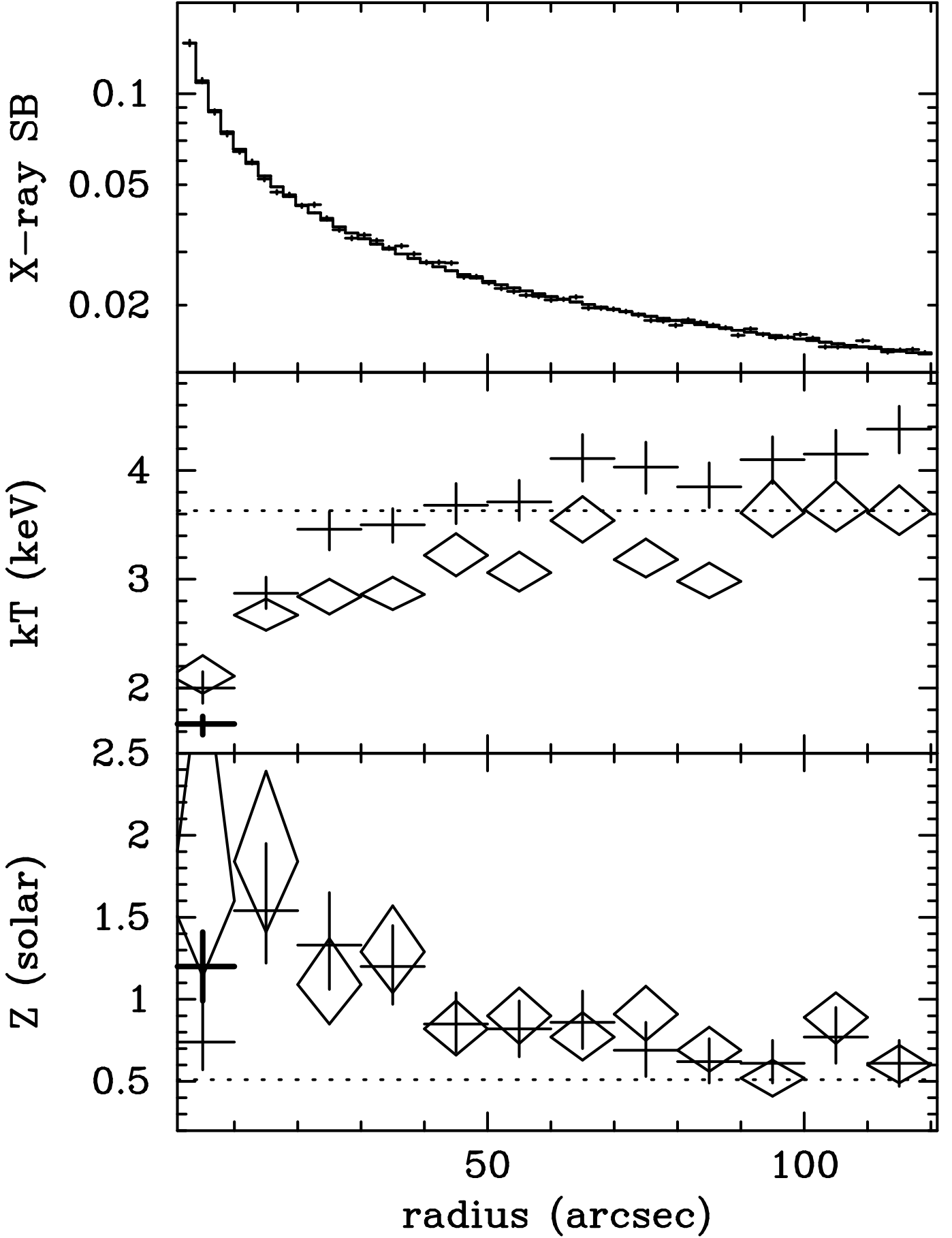}
\caption{\scriptsize Radial profiles of the surface brightness,
  temperature, and abundance from top to lower.  The top panel shows the
  data and a best-fit $\beta$ model. In the lower 2 panels, the crosses
  and diamonds show the results of 1--9 keV and 2--9 keV band
  fits. Errors are 90\% and 1-sigma confidence, respectively. The thick
  cross for the center only represents the results of a 2-temperature
  fit with fixed parameters for the projected ICM component. Each
  horizontal dotted line represents the ASCA result in the central
  $r=0'-5'$ region from \citet{ezawa97}. }
\label{rdists}
\end{center}
\end{figure}

%%%%%% Figure 7 %%%%%%%%%%

\begin{figure}[htb]
\begin{center}
%\plottwo{f7a.eps}{f7b.eps}
\caption{Temperature ({\it left}) and abundance ({\it right}) maps of
the central $2'\times2'$ region of AWM~7 obtained by spectral fits with a single temperature model. Each annular ring has a radial width of $10''$. The contours indicate the
smoothed 0.5--10 keV band image.}
\label{2dmaps}
\end{center}
\end{figure}

\end{document}